\documentclass{article}
\usepackage[utf8]{inputenc}

\title{Monte Carlo Simulation for Trading Under a L\'evy-Driven Mean-Reverting Framework}
\author{Tim Leung\thanks{Department of Applied Mathematics, University of Washington, Seattle, WA, USA. Email: \href{mailto:timleung@uw.edu}{timleung@uw.edu}.}
\and Kevin W. Lu\thanks{Research School of Finance, Actuarial Studies and Statistics, Australian National University, Canberra, ACT, Australia. Email:  \href{mailto:kevin.lu@anu.edu.au}{kevin.lu@anu.edu.au}.}}
\date{\today}

\usepackage{natbib}
\usepackage{graphicx}
\usepackage{amsmath,amsfonts,amssymb,amsthm}
\usepackage[colorlinks,citecolor=red,linkcolor=blue]{hyperref}

\numberwithin{equation}{section} 
\allowdisplaybreaks[4] 
\let\originalleft\left
\let\originalright\right
\def\left#1{\mathopen{}\originalleft#1}
\def\right#1{\originalright#1\mathclose{}}

\theoremstyle{definition}

\theoremstyle{remark}

\newcommand{\rmd}{{\rm d}}

\newcommand{\eqd}{\stackrel{D}{=}}

\newcommand{\EE}{\mathbb{E}}

\newcommand{\RR}{\mathbb{R}}

\newcommand{\PP}{\mathbb{P}}

\newcommand{\PPP}{{\cal P}}

\newcommand{\ZZZ}{{\cal Z}}

\newcommand{\eins}{{\bf 1}}

\newcommand{\bfnull}{{\bf 0}}

\newcommand{\bfb}{{\bf b}}
\newcommand{\bfc}{{\bf c}}

\newcommand{\bfx}{{\bf x}}

\newcommand{\bfd}{{\bf d}}

\newcommand{\bfB}{{\bf B}}
\newcommand{\bfC}{{\bf C}}

\newcommand{\bfT}{{\bf T}}
\newcommand{\bfU}{{\bf U}}
\newcommand{\bfV}{{\bf V}}

\newcommand{\bfX}{{\bf X}}
\newcommand{\bfY}{{\bf Y}}
\newcommand{\bfZ}{{\bf Z}}
\newcommand{\bfalpha}{\boldsymbol{\alpha}}
\newcommand{\bfbeta}{\boldsymbol{\beta}}

\newcommand{\bfepsilon}{\boldsymbol{\epsilon}}

\newcommand{\bfmu}{\boldsymbol{\mu}}

\newcommand{\bftheta}{\boldsymbol{\theta}}

\newcommand{\bfeta}{\boldsymbol{\eta}}

\newcommand{\myCov}{\operatorname{Cov}}
\newcommand{\myVar}{\operatorname{Var}}
\newcommand{\myCorr}{\operatorname{Corr}}

\newcommand{\wt}{\widetilde}
\newcommand{\tr}{\diamond}

\newcommand{\given}{{\,\vert\,}}

\newcommand{\wh}{\widehat}

\begin{document}

\maketitle
\begin{abstract}

We present a Monte Carlo approach to pairs trading on mean-reverting spreads modeled by L\'evy-driven Ornstein-Uhlenbeck processes. Specifically, we focus on using a variance gamma driving process, an infinite activity pure jump process to allow for more flexible models of the price spread than is available in the classical model. However, this generalization comes at the cost of not having analytic formulas, so we apply Monte Carlo methods to determine optimal trading levels and develop a variance reduction technique using control variates. Within this framework, we numerically examine how the optimal trading strategies are affected by the parameters of the model. In addition, we extend our method to bivariate spreads modeled using a weak variance alpha-gamma driving process, and explore the effect of correlation on these trades.

\vspace{0.5em} \noindent {\em Keywords:}  Pairs trading, Monte Carlo simulation, L\'evy process, Ornstein-Uhlenbeck process, mean reversion, variance gamma process.
       
\vspace{0.5em} \noindent {\em 2020 MSC Subject Classification:} 91G60, 65C05 (Primary) 60G51, 60G10, 62P05 (Secondary)
\end{abstract}
\newpage

\section{Introduction}

Many empirical studies have demonstrated examples of mean-reverting spreads in different markets. They can be generated by taking long and short positions in stocks and ETFs (\cite{gatev2006pairs,avellaneda2010statistical,huck2015pairs}), futures contracts (\citeauthor{futuresBS,futuresDaiKwok}), physical commodity and commodity ETFs (\cite{kanamura2010profit,meanReversionBook}), as well as cryptocurrencies (\cite{leung2019constructing}).

Another popular approach, the stochastic spread method captures the path behavior of the spread through a stochastic process with mean reversion. Given a mean-reverting spread process, we consider the problem of pairs trading, that is, extracting trading signals from the model based on the idea that a trade should be entered when the spread is far from the mean, and exited when it returns to the mean, so that the trader profits based on the tendency of the process to mean revert. The construction of spreads and extraction of trading signals are derived from the analysis of the underlying model. For instance, \cite{elliott2005pairs} propose a mean-reverting Gaussian Markov chain model to describe spread dynamics. The model's estimates are compared with observations of the spread to determine appropriate trading decisions. They solve an optimal double-stopping problem to analyze the timing of entry and exit subject to transaction costs. 

Most prominently, Ornstein-Uhlenbeck (OU) processes have been used to model spreads for pairs trading. For classical OU processes, which are driven by Brownian motion, closed-form solutions to the pairs trading problem were obtained in \cite{leung2015optimal,LiLP20}.

L\'evy processes have been widely applied in finance to model price processes with jumps (\cite{CT,sco}), which leads us to model price spreads with jumps. These L\'evy models often provide a better fit than models based on Brownian motion (\cite{MCC98, MiSz17, sco}), where the first two references focus on variance gamma-based models. Jumps can also arise from hard-to-borrow stocks (\cite{AvLi}) or illiquid stocks (\cite{GSS22}). This motivates going beyond the classical OU processes to consider L\'evy-driven Ornstein-Uhlenbeck processes (LDOUP) to model mean-reverting spreads with jumps in pairs trading.

LDOUPs been applied to other areas of mathematical finance such as modeling stochastic volatility (\cite{BNSh01}), energy prices and energy derivatives  (\cite{BeSc14,CKM18,Sab20}), credit risk  (\cite{CS092}), among others. LDOUPs also have increased flexibility for capturing skewness, kurtosis, and dependence in the multivariate setting. Parametric estimation of LDOUPs using maximum likelihood has been considered in the univariate and multivariate cases by \cite{vstt} and \cite{Lu21}, respectively, and these methods can be applied to fit LDOUPs to spreads. However, in the context of pairs trading, LDOUPs have been applied restricting the driving process to compound Poisson processes with double exponential jumps.   This setup is used since  \cite{BN08} derived an analytic formula for the expected exit time which relies on the memorylessness property of the exponential jumps  to deal with the overshoots.  Without this assumption,   the overshoots present a difficult technical hurdle to finding analytic formulas for the passage times of these processes in general, which requires Monte Carlo methods to overcome.  Based on \cite{BN08}, \cite{WZZ20} provide an analytic value function for pairs trading, while \cite{ES19}  use the analytic formula of \cite{BN08}, while ignoring overshoots, to conduct an empirical study of optimal pairs trading.

In this paper, we discuss a framework for pairs trading using L\'evy-driven Ornstein-Uhlenbeck processes to model the mean-reverting spread, which works with any LDOUP that can be simulated.  Specifically, we focus on a variance gamma driving process, which is a time-changed Brownian motion and a pure-jump, infinite activity L\'evy process with skewness and kurtosis parameters. By using this driving process, these properties extend to the LDOUP, allowing for more flexible models of the price spread than is available in the classical OU model, and in contrast to the finite activity processes considered previously. However, the generalization comes at the cost of not having analytic formulas. Hence, we develop a Monte Carlo method to evaluate the trading strategies and determine the optimal trading levels. In particular, we introduce an array of control variates to reduce the variance of the Monte Carlo estimator of the value function.  We derive a formula for the variance reduction ratio when the control variate is used to estimate the mean and variance of the profit simultaneously,   which extends the usual approach where only one parameter is estimated by control variates. Furthermore, in our numerical implementation, we examine how the parameters of the model affect the optimal trading levels and optimal expected profit in various scenarios. Specifically, we demonstrate numerically how skewness in the distribution of the spread leads to asymmetry in the optimal trading levels, how the optimal exit level can differ from the stationary mean when the process starts sufficiently far away from it and the discount rate is large, and it is otherwise difficult to find other situations where this occurs. And lastly, we demonstrate how the jumps can increase the optimal entry level and optimal expected profit.

We note that all the studies mentioned above focus on the trading performance of a single spread. The recent paper by \cite{LeeLeungNing2023} introduces a framework for trading multiple mean-reverting spreads simultaneously, where capital is dynamically allocated among different spreads based on their statistical characteristics. In this paper, we also consider the trading problem involving two mean-reverting spreads modeled by an LDOUP driven by a weak variance alpha-gamma process, which is a multivariate generalization of the variance gamma process introduced in \cite{BLM17a}. For bivariate pairs trading, we estimate the expected profit via Monte Carlo simulation and examine the effect of the correlation between spreads on the optimal trading levels and optimal expected profit. We demonstrate an example where the optimal strategy is to effectively pick the best spread mostly ignoring correlation, and an example where having low absolute correlation is more profitable when monitoring multiple spreads with marginal components that follow the same law.

The rest of the paper is structured as follows. 
In Section \ref{sect-LDOUP}, we review L\'evy-driven Ornstein-Uhlenbeck processes and required properties used in this study. The trading problem is formulated in Section \ref{sect-trading}. We discuss our Monte Carlo estimation framework in Section \ref{sect-MC} and present the numerical results in Section \ref{sect-numerics}. In Section \ref{sect-two-pairs}, we examine the trading problem involving two mean-reverting spreads. Concluding remarks are provided in Section \ref{sect-conclude}.

\section{L\'evy-Driven Ornstein-Uhlenbeck Process}\label{sect-LDOUP}

Let $\bfZ\sim L^n(\bfmu,\Sigma,\ZZZ)$ denote an $n$-dimensional L\'evy process with characteristic triplet $(\bfmu,\Sigma,\ZZZ)$, where $\bfmu\in\RR^n$, $\Sigma\in\RR^{n\times n}$ is a covariance matrix  and $\ZZZ$ is an $n$-dimensional  L\'evy measure. See \cite{bert96,sato99,sco} for references on L\'evy processes.

A \emph{L\'evy-driven Ornstein-Uhlenbeck process (LDOUP)} is defined by the stochastic
differential equation
\begin{align}
\rmd\bfX(t) =-\lambda\bfX(t)\rmd t + \rmd \bfZ(\lambda t),\quad \bfX(0)=\bfX_0,\quad t\geq 0,\label{sde}
\end{align}
where $\bfZ\sim L^n(\bfmu,\Sigma,\ZZZ)$ is the \emph{background driving L\'evy process (BDLP)}, $\lambda>0$, and $\bfX_0$ is a random vector independent of $\bfZ$. Note that there is no loss in generality in using the drift term $-\lambda\bfX(t)\rmd t$ instead of $\lambda(\overline{\bfmu}-\bfX(t))\rmd t$ for some $\overline{\bfmu}\in\RR^n$, since the stationary mean of $\bfX$ is controlled by $\EE[\bfZ(1)]$. Also, there is no loss in generality in using the BDLP $\bfZ\circ (\lambda I)$, where $I$ is the identity function, instead of $\bfZ$,  since any L\'evy process $\wt\bfZ\sim L^n(\wt\bfmu,\wt\Sigma,\wt\ZZZ)$ can be written in the form $\wt\bfZ\eqd\bfZ\circ (\lambda I)$, where $\bfZ\sim L^n(\wt\bfmu/\lambda,\wt\Sigma/\lambda,\wt\ZZZ/\lambda)$. However, the former is more convenient since it leads to a stationary distribution that does not depend on $\lambda$.  The exposition on LDOUP here follows \cite{Lu21}, while some additional references for LDOUPs include  \cite{mas04,saya,SaYa85}.

The solution of \eqref{sde} is
\begin{align}\label{ousoln}
\bfX(t)=e^{-\lambda t} \bfX(0) +e^{-\lambda t}\int_0^t e^{\lambda s}\,\rmd\bfZ(\lambda s),\quad t\geq 0.
\end{align}
For $t_0=0,t_1=\Delta , \dots,t_q= q\Delta$, at the observation time $t_i$, we have
\begin{align*}
\bfX(t_i)&=e^{-\lambda \Delta}\left( \bfX(t_{i-1}) +\int_{t_{i-1}}^{t_i} e^{\lambda s}\,\rmd\bfZ(\lambda s)\right),\quad i=1,\dots,q.
\end{align*}
The stochastic integral term is iid, and we let $\bfZ^*(\Delta)$ be the random vector with this distribution, so it is interpreted as the innovation term, up to a factor  $e^{-\lambda \Delta}$, of an AR(1) process. Specifically, we define
\begin{align*}
\bfZ^*(\Delta) = \int_0^\Delta e^{\lambda s}\,\rmd\bfZ(\lambda s) \eqd \int_{t_{i-1}}^{t_{i}}e^{\lambda s}\,\rmd\bfZ(\lambda s),\quad \Delta >0.
\end{align*}
The random vector $\bfZ^*(\Delta)$ is infinitely divisible, so it is determined by its characteristic exponent, which is given by
\begin{align}
\Psi_{\bfZ^*(\Delta)}(\bftheta) ={}& \int_{0}^{\lambda\Delta} \Psi_\bfZ (e^{ t}\bftheta)\,\rmd t,\quad \bftheta\in\RR^n, \label{pzizstar}
\end{align}
where $\Psi_{\bfZ}$ is the characteristic exponent of the BLDP $\bfZ$.
Combining these results, the LDOUP can be written as
\begin{align}
\bfX(t_i)&=e^{-\lambda \Delta}\left( \bfX(t_{i-1}) +\bfZ^*(\Delta)^{(i) }\right),\quad i=1,\dots,q,\label{xsim}
\end{align}
where $\bfZ^*(\Delta)^{(i)} $ is iid with the distribution of $\bfZ^*(\Delta)$.

Next, we consider specific examples of LDOUP processes where the BDLP is a tractable pure-jump infinite activity  L\'evy process to model price spreads for pairs trading. We consider both the univariate and multivariate cases, where the BDLP  is a variance gamma and a weak variance alpha-gamma process, respectively.

\subsection{Univariate LDOUP Driven by a Variance Gamma Process}

The variance gamma process was introduced in  \cite{MaSe90} and additional details can be found in \cite{MCC98}. Let $\Gamma_S(a,b)$ denote a gamma subordinator with shape $a>0$ and rate $b>0$, and $BM^n(\bfmu,\Sigma)$ denote a Brownian motion with drift $\bfmu=(\mu_1,\dots,\mu_n)\in\RR^n$ and covariance matrix $\Sigma=(\Sigma_{kl})\in\RR^{n\times n}$. A L\'evy process $\bfV\sim VG^n(b,\bfmu,\Sigma)$ is a \emph{variance gamma (VG) process} if
\begin{align*}
\bfV\eqd \bfeta I + \bfB\circ (G,\dots,G),
\end{align*}
where $\bfB\sim BM^n(\bfmu,\Sigma)$ and $G\sim\Gamma_S(b,b)$ are independent,  $\bfeta\in\RR^n$, and $I$ is the identity function. The parameter $\bfeta$ adds a drift.

We call the univariate LDOUP $X\sim OU\text{-}VG(\lambda,b,\mu,\sigma^2,\eta)$ an \emph{$OU\text{-}VG$ process} if  its BDLP is $Z\sim VG^1(b,\mu,\sigma^2,\eta)$.

\subsection{Multivariate LDOUP Driven by a Weak Variance Al\-pha-Gamma Process}\label{sect-MLDOUP}

The weak variance alpha-gamma (WVAG) process was introduced in \cite{BLM17a}.  Let $n\geq 2$, $a> 0$, $\bfalpha=(\alpha_1,\dots,\alpha_n) \in(0,1/a)^n$, $\beta_k:= (1-a\alpha_k)/{\alpha_k}$, $k=1,\dots,n$ and $\bfeta\in\RR^n$. Let $\bfalpha\tr\bfmu:= (\alpha_1\mu_1,\allowbreak\dots,\alpha_n\mu_n)\in\RR^n$ and $\bfalpha\tr\Sigma:=(\Sigma_{kl} (\alpha_k\allowbreak\wedge \alpha_l))\in\RR^{n\times n}$. We define  $\bfZ\sim WVAG^n(a,\bfalpha,\bfmu,\Sigma,\bfeta)$ as a \emph{weak variance alpha-gamma (WVAG)} process if
\begin{align}\label{wvagpropb}
\bfZ\eqd \bfeta I+ \bfV_0+(V_1,\dots,V_n),
\end{align}
where $\bfV_0\sim VG^n( a,a\bfalpha\tr\bfmu,a\bfalpha\tr\Sigma)$ and  $V_k\sim  VG^1(\beta_k,\allowbreak\alpha_k\beta_k\mu_k,\alpha_k\beta_k\Sigma_{kk})$, $k=1,\dots,n$, are independent. This formulation gives a direct method of simulating the WVAG process in terms of VG processes, and seeing that the process has common and idiosyncratic jumps.

The original definition of the WVAG process, 
\begin{align}
\bfZ \eqd \bfB\odot \bfT,\label{ws}
\end{align}
where $\bfB\sim BM^n(\bfmu,\Sigma)$, $\bfT$ an $n$-dimensional alpha-gamma subordinator, and $\odot$ is the weak subordination operation, gives a  multivariate time-change interpretation to the process, generalizing the univariate VG process. Indeed, the marginal components of the WVAG process are general univariate VG processes with $Z_k \sim VG^1(1/\alpha_k,\mu_k,\allowbreak\Sigma_{kk},\eta_k)$.

We call the multivariate LDOUP $\bfX\sim OU\text{-}WVAG^n(\lambda,a,\bfalpha,\bfmu,\Sigma,\bfeta)$  an  \emph{OU\text{-}WVAVG process} if its BDLP is $\bfZ\sim WVAG^n(a,\bfalpha,\bfmu,\Sigma,\bfeta)$.

Based on the moment formulas (see \cite{MiSz17}), the parameters of the WVAG process can be interpreted as follows. The marginal parameters $\alpha_k,\mu_k,\Sigma_{kk},\eta_k$ are kurtosis, skewness, variance, and location parameters, respectively, that affect the marginal component distribution, while $a,\Sigma_{ij}$, $i\neq j$, are joint parameters. In particular, as $\alpha_k\to0$ (equivalently, the VG parameter  $ b_k=1/\alpha_k\to\infty$ in the $k$th component), the marginal component of the WVAG process converges in law to Brownian motion so the sample paths become closer to continuous.

\subsection{Moment Formulas}

For a random vector $\bfU=(U_1,\dots,U_n)$, denote the mean by $m_1(\bfU):=\EE[\bfU]$, and the $k$th central moment of $\bfU$ by $m_k(\bfU):=(\EE[(U_1-\EE[U_1])^k],\dots,\EE[(U_n\allowbreak-\EE[U_n])^k])$, $k\geq2$.

For a general BDLP $\bfZ$,  \cite[Lemma 1]{Lu21} gives formulas for the moments of $\bfZ^*(\Delta)$ in terms of the moments of $\bfZ$.  Combining this with \eqref{ousoln} and conditional on $\bfX_0$ (or assuming it is constant), the moments of the LDOUP $\bfX$ are given by
\begin{align*}
m_1(\bfX(t)) &= e^{-\lambda t}\bfX_0 +(1-e^{-\lambda t})m_1(\bfZ(1)),\nonumber\\
m_2(\bfX(t))  &= \left( \frac{1-e^{-2\lambda t}}{2}\right)m_2(\bfZ(1)),\\
m_3(\bfX(t))  &= \left( \frac{1-e^{-3\lambda t}}{3}\right)m_3(\bfZ(1)),\\
\myCov(X_k(t),X_l(t))&=\left(\frac{1-e^{-2\lambda t}}{2}\right)  \myCov(Z_k(1),Z_l(1)),\quad k\neq l.\nonumber
\end{align*}
The stationary distribution $\bfY$ of $\bfX$ is the distribution such that if $\bfX_{0}\eqd \bfY$, then $\bfX(t)\eqd\bfY$ for all $t\geq0$. Thus, the stationary mean, which is the mean of the stationary distribution, is $\overline{\bfmu} =\EE[\bfZ(1)]$. This is the level that we expect the process to return to in pairs trading.

Now specializing to the case where $\bfX\sim OU\text{-}WVAG^n(\lambda,a,\bfalpha,\bfmu,\Sigma,\bfeta)$,  and using the moment formulas for the WVAG process in \cite[Remark 4 and Appdendix A.1]{MiSz17}, we have

\begin{align}
\EE[X_1(t)] &= e^{-\lambda t}X_0 +(1-e^{-\lambda t})(\eta_1+\mu_1), \label{cv1}\\
\myVar(X_1(t)) &= \left(\frac{1-e^{-2\lambda t}}{2}\right){(\Sigma_{11}+\alpha_1\mu_1^2)}, \label{cv2}\\
\operatorname{Skew}(X_1(t)) &= \frac{(\frac{1-e^{-3\lambda t}}{3})(3\Sigma_{11}\mu_1\alpha_1 +2\mu^3_1\alpha_1^2)}{\left((\frac{1-e^{-2\lambda t}}{2})(\Sigma_{11}+\mu_1^2\alpha_1)\right)^{3/2}}\label{sk1}\\
\myCov(X_1(t),X_2(t))&=\left(\frac{1-e^{-2\lambda t}}{2}\right) a((\alpha_1\wedge \alpha_2)\Sigma_{12}+\alpha_1\alpha_2\mu_1\mu_2).\label{covx}
\end{align}
Furthermore, $\overline{\bfmu} = \bfmu+\bfeta$. These formulas will be used for control variates in the Monte Carlo method. Note that \eqref{cv1} and \eqref{cv2} apply to OU-VG processes with $b_1=1/\alpha_1$.

\section{Pairs Trading Problem}\label{sect-trading}

Suppose the spread of a pair of assets or the price of a portfolio of assets is mean-reverting and follows a univariate LDOUP $X$.

Now we proceed to determine the levels at which trades should be entered or exited. Consider the following pairs trading strategy on the spread $X$: Enter a short position in $X$ when it is above $\overline{\mu}+d$, and close the position by buying when it is below $\overline{\mu}+c$, where $0\leq c<d$. Enter a long position when it is below  $\overline{\mu}-d$, and close the position by selling when it is above $\overline{\mu}-c$. Fix a terminal time $T>0$, where the trade is closed by time $T$ if it has not been closed already. Thus, $d$ and $c$ are the entry and exit levels relative to the mean reversion level $\overline{\mu}$.

\begin{figure}[!htb]
\begin{center}
    \includegraphics[width=.8\textwidth]{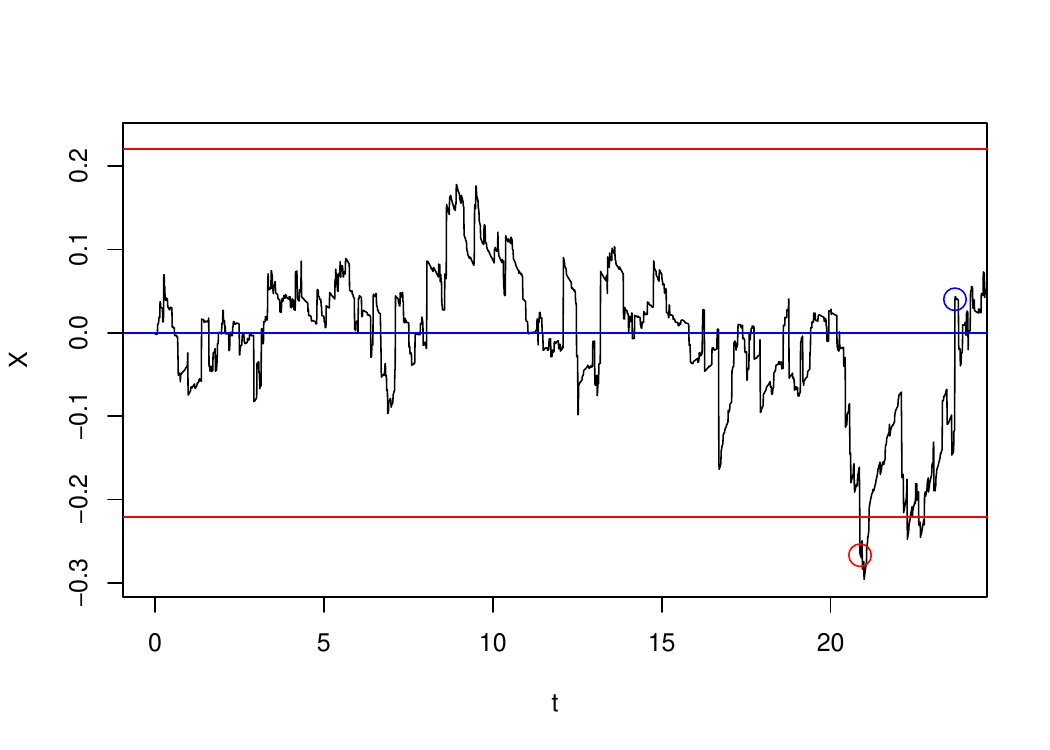}
    
    \caption{For the spread process $X$ (black line), a trade is entered when the price passes $\pm d$, whichever happens first (red lines), and exited when it passes $\pm c=0$  (blue line). The value of the spread at the entry (red point) and exit (blue point) times are shown. Here, $X\sim OU\text{-}VG(1,5,0,0.015,0)$.}\label{fig:trade}
\end{center}
\end{figure}

This trading strategy is depicted for a sample path of $X$ in Figure \ref{fig:trade}. Note that the price at which the trade enters and exits is not the same as the entry and exit levels due to the presence of jumps. In this model, there is always some degree of overshoot.

More generally, we can consider possibly different entry levels $d_+$, $d_-$ from above and below $\overline{\mu}$, and their respective exit levels $c_+, c_-$. These satisfy  $0\leq c_{\pm}<d_{\pm}$.

The relevant passage times are
\begin{align*}
\tau_{d_+}&=\inf\{t\geq 0 : X(t)>\overline \mu + d_+\}\wedge T,\\
\tau_{c_+}&=\inf\{t\geq \tau_{d_+} : X(t)< \overline \mu + c_+\}\wedge T,\\
\tau_{d_-}&=\inf\{t\geq 0 : X(t)< \overline \mu - d_-\}\wedge T,\\
\tau_{c_-}&=\inf\{t\geq \tau_{d_-} : X(t)> \overline \mu - c_-\}\wedge T. 
\end{align*}
Let $r\in\RR$ be a discount rate, then $P$ is the discounted profit over one trade cycle given by
\begin{align}
\begin{split}
P:=P(c_+,c_-,d_+,d_-) = {}&\eins_{\{\tau_{d_+} \leq \tau_{d_-}\}} e^{-r\tau_{c_+} }   (X(\tau_{d_+}) - X(\tau_{c_+})) \\
&\quad+\eins_{\{\tau_{d_+}  > \tau_{d_-}\}} e^{-r\tau_{c_-} }   (X(\tau_{c_-})-X(\tau_{d_-})  ).\label{profit}
\end{split}
\end{align}
For convenience, we will refer to this simply as the profit.
The value function is 
\begin{align}
V:=V(c_+,c_-,d_+,d_-)&=\EE[P]-\gamma\myVar(P)\nonumber\\
& = \EE[P]-\gamma\EE[P^2] +\gamma\EE[P]^2,\label{objfn}
\end{align}
where $\gamma\geq 0$ is a parameter which penalizes trading strategies with high variance. If $\gamma=0$, the value function is the expected profit.

We want to find the optimal levels $c_{\pm},d_{\pm}$ for the pairs trading strategy to maximize the value function, that is,
\begin{align*}
\max_{c_+,c_-,d_+,d_-} V(c_+,c_-,d_+,d_-).
\end{align*}

We develop a Monte Carlo approach to simulate sample paths of the LDOUP $X$, to estimate $V(c_+,c_-,d_+,d_-)$ on $0\leq c_{\pm}<d_{\pm}$, and in turn numerically solve the maximization problem for the optimal trading levels. Given the use of simulation, we also discuss control variates for variance reduction.

\section{Monte Carlo Method}\label{sect-MC}

Let $m$ be the number of Monte Carlo simulations. For a spread process following a univariate LDOUP $X$, suppose we have $m$  simulations of $X(t_0),\dots,X(t_q)$  with a small stepsize $\Delta>0$. Then we can compute the random variable $P$,  and hence obtain the Monte Carlo estimate of $\EE[P]$ and $\EE[P^2]$ to estimate the value function $V$ in \eqref{objfn}. The same sample paths are used for all evaluations of the same value function. While the Monte Carlo method presented here works for any LDOUP that can be fitted and simulated, we specifically focus on the OU-VG process.

\subsection{Simulating the Spread}

Let the spread process be $X\sim OU\text{-}VG(\lambda,b,\mu,\sigma^2,\eta)$. We assume that the parameters are known. For example, they may be estimated using maximum likelihood as in \cite{vstt,Lu21}.

It is possible to exactly simulate $Z^*(\Delta)$ for the OU-VG process, and hence $X(t_0),\allowbreak,\dots,X(t_q)$  by \eqref{xsim}, using the method of \cite{Sab20,QDZ19}.  For $G\sim \Gamma_S(a,b)$, define the random variable $G^*(\Delta) \eqd e^{\lambda\Delta}(\wt G + \wt C)$, where $\wt G\sim \Gamma(a\lambda\Delta, be^{\lambda\Delta})$ is gamma where the parameters are the shape and rate, respectively, and $\wt C\sim  CP(a\lambda^2\Delta^2/2,\allowbreak \PPP_J)$ is compound Poisson where the parameters are the rate and jump distribution, respectively. Here, $\PPP_J$ is the probability law of $J$ such that $J\given U \sim \operatorname{Exponential}(be^{\lambda\Delta \sqrt{U}})$ where the parameter is the rate, and $U\sim \operatorname{Uniform}(0,1)$. Then
\begin{align*}
Z^*(\Delta)\eqd \eta(e^{\lambda \Delta}-1)+G_+^*(\Delta) -   G_-^*(\Delta),
\end{align*}
where $G_+\sim \Gamma_S(b,b_+)$, $G_-\sim \Gamma_S(b,b_-)$, and 
\begin{align*}
b_{\pm} := \frac{2b}{\sqrt{\mu^2 + 2\sigma^2b}\pm\mu}.
\end{align*}

\subsection{Control Variates} \label{cv-sec}

Next, we use control variates to reduce the variance of the Monte Carlo estimators of the value function. We refer to  \cite[Chapter 4]{Glas} for the details of the method. Specifically, to estimate $\EE[P]$, the method of control variates defines the random variable  $P_{C} = P - \beta(C-\EE[C])$, where $C$ is the control variate for which $\EE[C]$ has an analytic representation, and $\beta$ is chosen to minimize $\myVar(P_{C})$, resulting in an unbiased estimator with reduced variance. Since the true value of $\beta$ is unknown, the control variate estimator $\wh P_{C}$ is the sample mean of $P_C$ with $\beta$ estimated by $\wh \beta$, the least squares estimator (LSE) of the slope in a linear regression where $C$ is the regressor and $P$ is the response variable. Note that $\wh \beta$ is also the sample estimate of the true  $\beta$.

This method can be extended to multiple control variates, and the control variate estimator is equivalent to the prediction of a linear regression. See Appendix \ref{appendix} for details.

We apply the control variate method to estimate both $\EE[P]$ and $\EE[P^2]$ in the value function \eqref{objfn}. Consider the control variates  $X(t_1)-\overline{\mu},\dots,X(t_p)-\overline{\mu}$, where $0,t_1,\dots, t_p$ are equally spaced points on $[0,T]$. When estimating $\EE[P^2]$, we include the additional control variates $(X(t_1)-\overline{\mu})^2,\dots,\allowbreak (X(t_p)-\overline{\mu})^2$. Note that $\EE[X(t)-\overline{\mu}]$ and $\EE[(X(t)-\overline{\mu})^2]$ can be analytically computed using \eqref{cv1} and \eqref{cv2}.

Further, we may want to consider the event of entering and exiting a trade, and the event of entering and not exiting over the observations $X(t_1),\dots, X(t_p)$ as additional control variates to approximate the indicator variable appearing in the profit function \eqref{profit}. But the event that $X(t_1),\dots, X(t_p)$ passes some level cannot be computed analytically for use as a control variate. Instead, the probability of a related event using the iid sequence of innovations $Z^*(\Delta)^{(i)}$, $i=1,\dots, p$ can be. Specifically, this motivates the inclusion of two additional control variates $\eins_A$ and  $\eins_B$. Define $A$ as the event that $Z^*(\Delta)^{(1)},\dots,Z^*(\Delta)^{(p)}$ ``enters a trade" (even though there is no actual trading on this sequence) by passing the level $e^{\lambda \Delta}(\overline\mu+d) -\overline\mu$ or $e^{\lambda \Delta}(\overline\mu-d) -\overline\mu$ and then ``exits the trade" by passing the level  $e^{\lambda \Delta}(\overline\mu+c) -\overline\mu$ or  $e^{\lambda \Delta}(\overline\mu-c) -\overline\mu$, respectively. 

The modified levels for $Z^*(\Delta)^{(i)}$ are  the levels such that
\begin{align*}
X(t_i) = e^{-\lambda \Delta}(X(t_{i-1})+Z^*(\Delta)^{(i)})
\end{align*}
passes the entry levels $\mu\pm d$ when  $X(t_{i-1})=\overline\mu$,  and similarly for the exit levels.  Further, $B$ is the same as $A$ except without exit.  Thus, $A$ and $B$ are the events that $Z^*(\Delta)^{(i)}$, $i=1,\dots,p$, enters and exits these modified levels, and enters and does not exit, respectively. Due to collinearity, the complementary event that the sequence does not enter is not included, and if $p=1$, the event $A$, which has probability 0, is also not included. We have expressed this for symmetric levels $c=c_+=c_-$ and $d=d^+=d_-$, but this can be modified in an obvious way for the asymmetric case.

Since  $Z^*(\Delta)^{(i)}$, $i=1,\dots,p$, are iid, we can compute the probabilities of $A$ and $B$ analytically since we can compute $\PP(Z^*(\Delta)\leq x)$ analytically by applying the Fourier inversion to the characteristic function of $Z^*(\Delta)$ determined by \eqref{pzizstar}. The Fourier inversion methodology is described in \cite[Section 4.1]{MiSz17}.

\subsection{Variance Reduction}

Consider the two separate linear regression models,
\begin{align}
    \bfY_1 = X_1\bfbeta_1 +\bfepsilon_1,\quad \bfY_2 = X_2\bfbeta_2 +\bfepsilon_2,\label{lr}
\end{align}
where $\bfY_1$ and $\bfY_2$ are the vector of $m$ simulated values of $P$ and $P^2$, respectively. In this section only, we let $X_1\in\RR^{(p_1+1)\times m}$ and $X_2\in\RR^{(p_2+1)\times m}$ be the design matrices, including an intercept.  As outlined in Section \ref{cv-sec}, $X_1$ includes the control variates $X(t_i)-\overline{\mu}$, $i=1,\dots,p$, $\eins_A$, $\eins_B$, and $X_2$ includes the control variates in $X_1$ in addition to  $(X(t_i)-\overline{\mu})^2$, $i=1,\dots,p$. Since each simulated sample path of the LDOUP is independent, we have $\myCov(\bfepsilon_i) = \Sigma_{\bfepsilon,ii} I$, $i = 1,2$ and $\myCov(\bfepsilon_1,\bfepsilon_2) = \Sigma_{\bfepsilon,12} I$ for some $2\times 2$ covariance matrix  $\Sigma_{\bfepsilon} = ( \Sigma_{\bfepsilon,ij} )_{i,j=1}^2$, and where $I$ is the identity matrix.

For each of the linear regression models, $i=1,2$, let  $\bfmu_{C,i}$ be the mean vector of the control variates computed as outlined in Section \ref{cv-sec}, then the predictions  $\bfx_i =(1,\bfmu_{C,i})$  are $\wh Y_i= \bfx_i '\wh\bfbeta_i $. As explained in Appendix \ref{appendix}, the control variate estimators of $\EE[P]$ and $\EE[P^2]$ are the predictions $\wh Y_1$ and  $\wh Y_2$, respectively. Thus, for each argument, the control variate estimator of the value function in \eqref{objfn} is 
\begin{align*}
    \wh V_{C} :=\wh Y_1-\gamma \wh Y_2 +\gamma \wh Y_1^2 = \mathbf{b} '\wh {\bfY} + \wh {\bfY}' A \wh {\bfY},
\end{align*}
 where
\begin{align*}
    A = \begin{pmatrix}
        \gamma & 0 \\ 0 & 0
    \end{pmatrix},\quad 
    \bfb = \begin{pmatrix}
        1\\-\gamma
    \end{pmatrix},
    \quad  
    \wh \bfY = \begin{pmatrix}
        \wh Y_1\\ \wh Y_2
    \end{pmatrix}.
\end{align*}
Let $\overline Y_1,\overline Y_2$ be the sample mean of $\bfY_1,\bfY_2$ respectively, then $\wh V_{{M}} := \overline Y_1-\gamma \overline Y_2 +\gamma \overline Y_1^2$ is the corresponding Monte Carlo estimator of the value function. 

Given that our approach involves parameters that are estimated by a function of two control variate estimators, rather than the usual situation of only one, we now derive the variance reduction factor in this context. For a large number of Monte Carlo simulations $m$, the asymptotic normality of the LSE gives $\wh\bfY\sim N(\bfmu_{C},\Sigma_{C})$ approximately for some $\bfmu_{C}\in\RR^2$ and $\Sigma_{C}\in\RR^{2\times 2}$. Using results on  multivariate linear regression, and the law of total variance, noting that $X_1$ and $X_2$ are random,  we have
\begin{align*}
    \bfmu_{C} &= \begin{pmatrix}
        \bfx_1' \bfbeta_1 \\ \bfx_2' \bfbeta_2
    \end{pmatrix},\\
    \Sigma_{C} &= \begin{pmatrix}
        \Sigma_{\bfepsilon,11}\bfx_1' \EE[(X_1'X_1)^{-1}]\bfx_1 &  \Sigma_{\bfepsilon,12}\bfx_1' \EE[(X_1'X_1)^{-1}X_1'X_2 (X_2'X_2)^{-1}]\bfx_2\\
        * &   \Sigma_{\bfepsilon,22}\bfx_2' \EE[(X_2'X_2)^{-1}]\bfx_2
    \end{pmatrix},
\end{align*}
 where $*$ denotes the rest of the matrix is completed by symmetry.
By the plug-in principle, these parameters are estimated using
\begin{align*}
    \wh\bfmu_{C} &= \begin{pmatrix}
        \bfx_1' \wh\bfbeta_1 \\ \bfx_2' \wh\bfbeta_2
    \end{pmatrix},\\
    \wh\Sigma_{C} &= \begin{pmatrix}
        \wh\Sigma_{\bfepsilon,11}\bfx_1' (X_1'X_1)^{-1}\bfx_1 &  \wh\Sigma_{\bfepsilon,12}\bfx_1' (X_1'X_1)^{-1}X_1'X_2 (X_2'X_2)^{-1}\bfx_2\\
        * &   \wh \Sigma_{\epsilon,22}\bfx_2'  (X_2'X_2)^{-1} \bfx_2
    \end{pmatrix},
\end{align*}
where $\Sigma_{\bfepsilon,ij}$ is estimated using 
\begin{align*}
    \wh \Sigma_{\bfepsilon,ij} = \frac{\wh\bfepsilon_i ' \wh\bfepsilon_j}{m-r-1},
\end{align*}
and $\wh\bfepsilon_1,\wh\bfepsilon_2$ are the residuals from the linear regressions in \eqref{lr}, $r=p_1$ for $i=j=1$, and $r=p_2$ otherwise.

Now since $\wh V_{C}$ is a quadratic form in  $\wh\bfY$, and $\wh\bfY\sim N(\bfmu_{C},\Sigma_{C})$ holds approximately for large $m$, by \cite[Theorem 5.2c--d]{ReSc08}, the variance of the  control variate estimator is
\begin{align}
    \myVar(\wh V_{C}) &\approx \bfb'\Sigma_{C} \bfb + 2 \operatorname{tr}((A\Sigma_{C})^2) + 4\bfmu' A\Sigma_{C} A \bfmu_{C} + 4 \bfb' \Sigma_{C} A\bfmu_{C} \label{cvvarest}
    \\&=(1+ 4\mu_{{C},1}\gamma+ 4(\mu_{{C},1})^2\gamma^2)\Sigma_{{C},11} + 2\gamma^2(\Sigma_{{C},11})^2 \nonumber\\
    &\quad -2\gamma(1+2\mu_{{C},1}\gamma)\Sigma_{{C},12} + \gamma^2\Sigma_{{C},22} \nonumber.
\end{align}
Then estimated variance $\wh\myVar(\wh V_{C})$ is obtained by replacing $\bfmu_{C}$ and $\Sigma_{C}$ with  $\wh\bfmu_{C}$ and $\wh\Sigma_{C}$, respectively. By the asymptotic normality of $\wh V_{{M}}$, \eqref{cvvarest} is also used to compute $\wh\myVar(   \wh V_{{M}})$ by replacing  $\wh\bfmu_{C}$ and $\wh\Sigma_{C}$ with $\wh\bfmu_M = (\overline{Y}_1,\overline{Y}_2)$ and $\wh\Sigma_M = \wh\Sigma_{Y}/m$, respectively,  where $\wh\Sigma_{Y}\in\RR^{2\times2}$ with the sample covariance $\wh{\myCov}(\bfY_i,\bfY_j)$ as its $(i,j)$-entry.

Thus, the estimated variance reduction factor of using control variates relative to Monte Carlo without control variates is
\begin{align*}
    R=\frac{ \wh\myVar(   \wh V_{C}) }{ \wh\myVar(   \wh V_{{M}}) }.
\end{align*}

In the context of a single parameter estimated using control variates, while the true variance reduction factor using the true value of $\bfbeta_i$ is no greater than 1, when using LSE $\wh\bfbeta_i$ instead, $R>1$ is possible. This topic is discussed in \cite[Section 4.1.3]{Glas}.

\section{Numerical Implementation and Analysis}\label{sect-numerics}

In this section, we implement the Monte Carlo method outlined above for various scenarios. The spread $X\sim OU\text{-}VG(\lambda,b,\mu,\sigma^2,\eta)$ is simulated  with stepsize $\Delta= 0.01$ and terminal time $T=50$.  In all cases, $\mu=-\eta$, and any changes to $\mu$ adjusts $\eta$ accordingly to ensure that the stationary mean is $\overline{\mu}=0$. Unless otherwise stated, the initial value is $X_0=0$, $\gamma=0$, the discount rate is $r=0.01$, and it is assumed for simplicity that the entry level  $d=d_+=d_-$ is symmetric and the exit level is fixed at $c=c_+ = c_-=0$. Thus, we optimize for the entry level $d$.  All results use $m= 10,000$ Monte Carlo simulations.

\subsection{Effect of Control Variates}   
Consider two examples using control variates.

\paragraph{Example 1.}  Suppose the spread $X$ has parameters $(\lambda,b,\mu,\sigma^2,\eta)=(1,1,-0.5,\allowbreak 0.015,0.5)$, and $\gamma=0.1$. 

\begin{figure}[!htb]
\begin{center}
    \includegraphics[width=.8\textwidth]{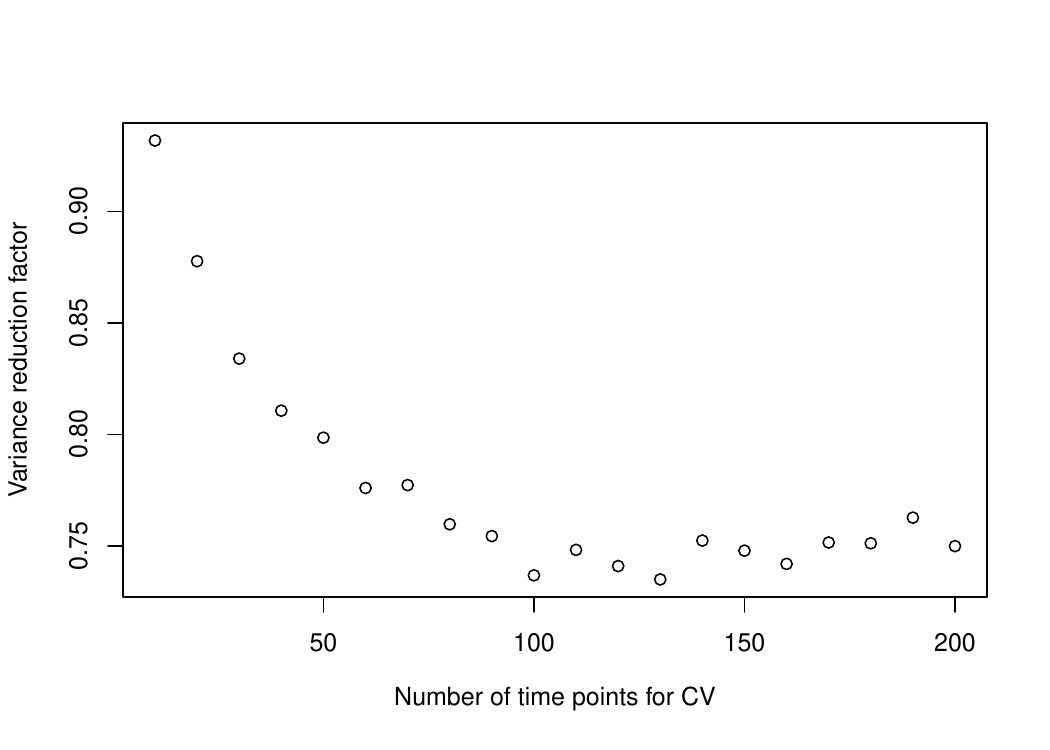}
    \caption{Estimated variance reduction factor $R$ as a function of the number of time points for the control variates $p$. The parameters are $(\lambda,b,\mu,\sigma^2,\eta)=(1,1,-0.5,0.015,0.5)$, $X_0=0$, $\gamma =0.1$, $r=0.01$.}\label{pcv}
\end{center}
\end{figure}

The number of time points $p$ used in the control variate method can be chosen to obtain the optimal variance reduction factor, however, this also depends on the entry level $d$. We optimize this iteratively by using no control variates to optimize $d$, then use this value of $d$ to optimize $p$ with the resulting optimal value denoted $p^*$. Doing this, Figure \ref{pcv} shows that for $d= 1.086$, the optimal number of time points for the control variates is $p^*=130$ when searching over $p=10, 20, \dots, 200$. This gives a moderate optimal variance reduction factor of $R^*=0.735$.

\begin{figure}[!htb]
\begin{center}
    \includegraphics[width=.8\textwidth]{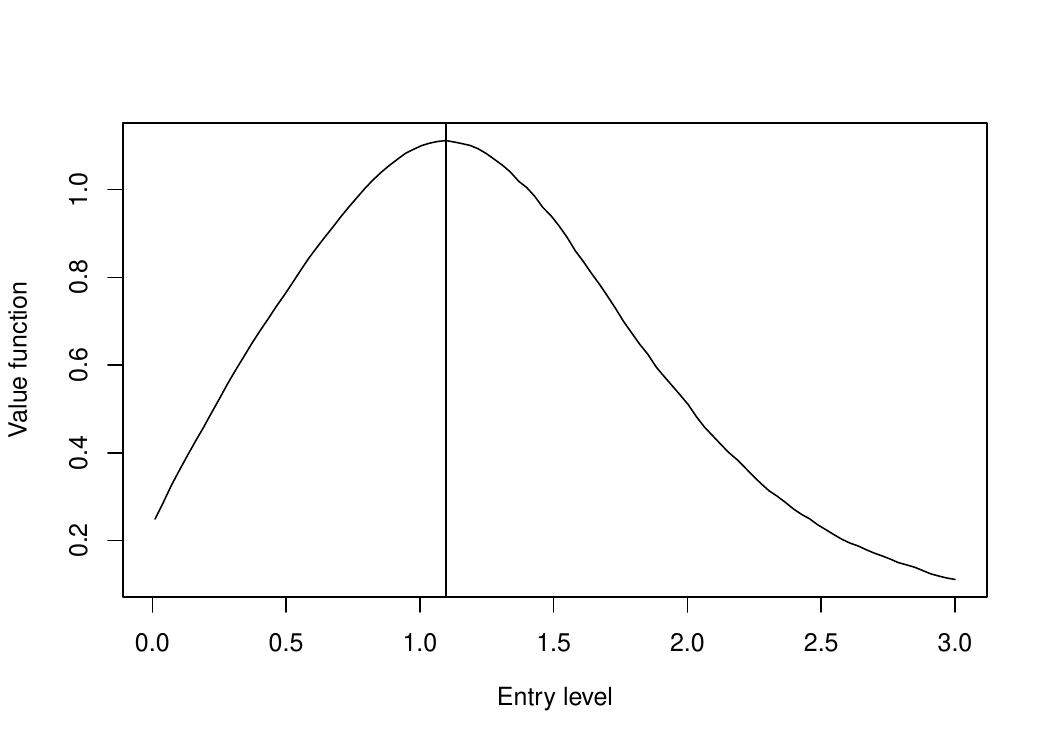}
    \caption{The estimate $\wh V_C (d)$ of the value function $V(d)$ and  the optimal entry level $d^*= 1.086$. The parameters are $(\lambda,b,\mu,\sigma^2,\eta)=(1,1,-0.5,0.015,0.5)$, $X_0=0$, $\gamma =0.1$, $r=0.01$.}\label{fig-valfn}
\end{center}
\end{figure}

Using $p^*=130$, Figure \ref{fig-valfn} shows the estimate of the value function  $V(d)$, which has optimal entry level $d^*= 1.086$. The control variate reduces $\myVar(\wh V_C)$, however, it does not change the optimal entry level $d^*$ in this example because the variability is already quite low, in fact, the coefficient of variation of $\wh V_C(d^*)$, the estimate of $V(d^*)$ conditional on $d^*$, is 0.0038 after control variates.

Different choices in the parameters result in different variance reductions. In Table \ref{cv-table}, results are shown varying $b,\mu$ (which also affects $\eta$), and for the parameters, decreasing $b>0$ increases  $R^*$, while decreasing $\mu<0$ reduces and then increases  $R^*$.

\begin{table}[htp]
\begin{center}
    \begin{tabular}{cc|cc}
        \hline 
        $b$ & $\mu$  & {$p^*$} & {$R^*$}  \\ \hline\hline
        $3$   & $-0.05$   & 70  &   0.951 \\
        $2$   & $-0.05$   & 120  & 0.904 \\
        $1$   & $-0.05$   & 100  & 0.871 \\
        $1$   & $-0.5$   & 130  &  0.735 \\
        $1$   & $-1$   & 110  & 0.826 \\
        \hline
    \end{tabular}
    \caption{For various values of $(b,\mu)$, the optimal number of time points for the control variates $p^*$ (searching over $p=10,20\dots,200$) and the optimal variance reduction factor $R^*$ are shown. The other parameters are $(\lambda,\sigma^2)=(1,0.015)$, $X_0=0$, $\gamma =0.1$, $r=0.01$.}\label{cv-table}
\end{center}
\end{table}

Consider the case shown in the second row of Table  \ref{cv-table} where $(b,\mu) =(2,-0.05)$ and $p^*=120$. Surprisingly, for large $\gamma$, such as $\gamma=1.5$, it is possible that the control variate method reduces the variance of the estimator of $\EE[P]$ with an estimated variance reduction factor  0.910, and $\EE[P^2]$ with estimated variance reduction factor    0.816, but increases the variance of the estimator of $V(d^*)=\EE[P]-\gamma\EE[P^2] +\gamma\EE[P]^2$ with variance reduction factor    1.139. In comparison, when $\gamma=0.1$, this does not occur, and those estimated variance reduction factors are 0.892, 0.793, 0.904, respectively.

\begin{figure}[!htb]
    \begin{center}
        \includegraphics[width=.8\textwidth]{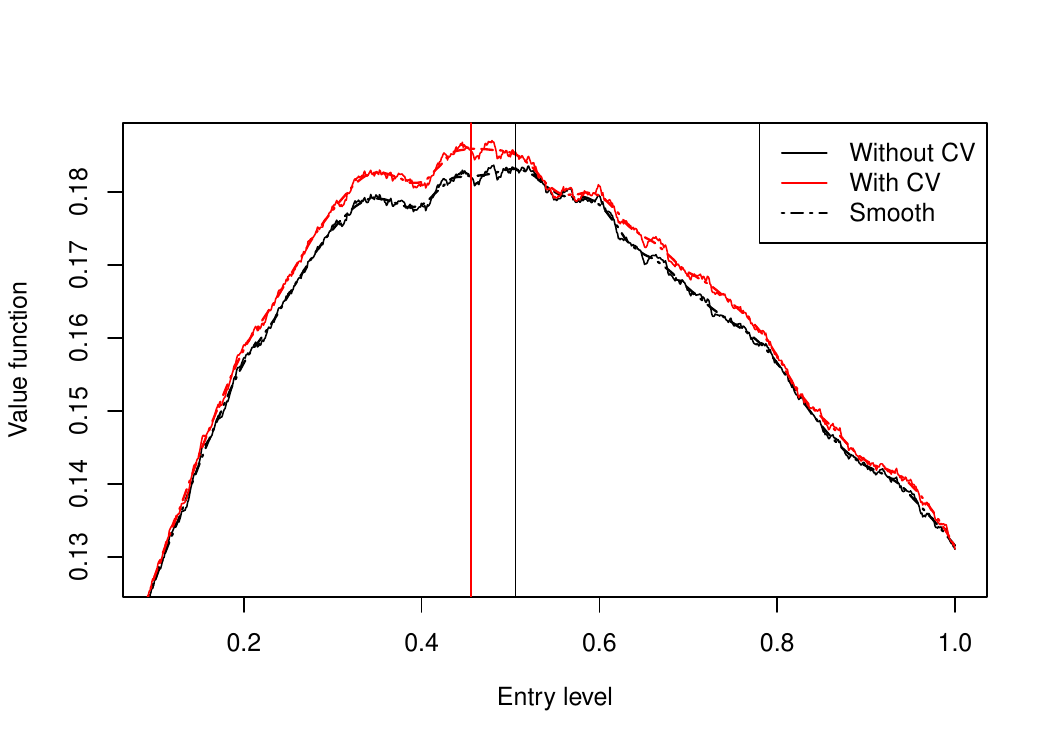} 
        \caption{The estimate of the value function $V(d)$ (solid lines) and the corresponding loess smooth (dashed lines) with (red lines) and without (black lines)  control variates. The optimal entry level is $d^*=0.506$ without using control variates, and $d^*=0.456$ using the optimal number of control variates $p^*=1$. The parameters are $(\lambda,b,\mu,\sigma^2,\eta)=(0.01,50,0.5,4,-0.5)$, $X_0=0$, $\gamma =0.1$, $r=0.01$.}\label{exp2plot}
    \end{center}
\end{figure}

\paragraph{Example 2.} We use the same setup as Example 1, except $X$ now has parameters $(\lambda,b,\mu,\sigma^2,\eta)=(0.01,50,0.5,4,-0.5)$. Since $\lambda\approx 0$, this process has little mean reversion, and is close in law to a variance gamma process, so the value function is estimated with a relatively large amount of simulation error as shown in Figure \ref{exp2plot}. To address this, a locally estimated scatterplot smoothing (loess) smoother is applied to the estimated value function before maximizing it. Then the optimal entry level is $d^*=0.506$ without using control variates, and $d^*=0.456$ using the optimal number of control variates, which is $p^*=1$. The optimal variance reduction factor is $R^*=0.817$ and the coefficient of variation of $\wh V_C(d^*)$ is  0.0281, which is consistent with the much larger variability of the estimated value function compared to Example 1, Figure \ref{fig-valfn}. Unlike in  Figure \ref{fig-valfn}, using control variates here has an effect on the estimate of $d^*$,  the size of this effect varies with the simulation. Note that $R^*$ is the variance reduction of estimating $V(d)$ for fixed $d$  using control variates, and the method is not designed to reduce the standard error of $d^*$, although reducing the former should be expected to reduce the latter.  For these optimal values,  only 63.91\% of simulations resulted in a trade that is entered and exited before the terminal time due to the slow mean reversion, compared to 90.88\% in Figure \ref{fig-valfn}.

\subsection{Asymmetry of Optimal Trading Levels}

From now on, we set $\gamma=0$ and do not include control variates. In the previous examples, we fixed $d=d_{+}=d_{-}$ for simplicity. The parameter $\mu$ controls the asymmetry of the stationary distribution $Y$ of the LDOUP $X$. If $\mu=0$, then the stationary distribution is symmetric and hence the optimal levels are equal, $d^*_+= d^*_-$. However, under asymmetry, we may want to optimize both $d_+$ and $d_{-}$ separately.

Here,  we consider the optimal level in four cases: from a large negative skew to no skew. The other parameters of $X$ are $(\lambda,b,\sigma^2)=(1,5,0.015)$.  The results are shown in Table \ref{asym}, where $\operatorname{Skew}(Y)$ is the skewness of the stationary distribution, which by \eqref{sk1}, is
\begin{align*}
    \operatorname{Skew}(Y) = \frac{2^{3/2}}{3}\left(\frac{3\sigma^2\mu/b +2\mu^3/b^2}{(\sigma^2 + \mu^2/b)^{3/2}}\right).
\end{align*}

\begin{table}[htb] 
\begin{center}
    \begin{tabular}{cc|cc}
        \hline 
        $\mu$ & $\operatorname{Skew}(Y)$ & {$d_+^*$} & {$d_-^*$}  \\ \hline\hline
        $-0.5$  & $-0.825$ &  0.362 & 0.488  \\
        $-0.2$ &  $-0.660$ & 0.231 & 0.293 \\
        $-0.05$ &  $-0.225$ & 0.213  & 0.227 \\
        $0$   & 0 &0.220   & 0.220 \\
        \hline
    \end{tabular}
    \caption{For various value of $\mu$, the optimal entry levels $d_+^*$ and $d_-^*$ are shown. The other parameters are $(\lambda,b,\sigma^2)=(1,5,0.015)$, $X_0=0$, $\gamma =0$, $r=0.01$.}\label{asym}
\end{center}
\end{table}

\begin{figure}[!htb]
\begin{center}
    \includegraphics[width=.8\textwidth]{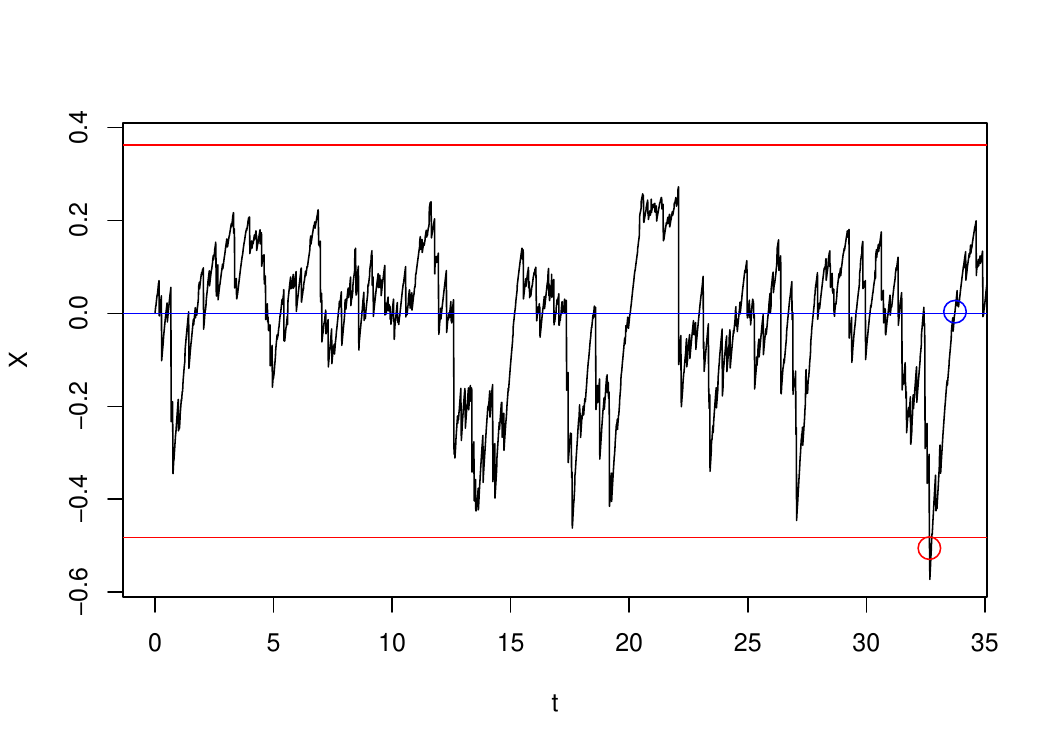}
        \caption{A sample path of the spread  $X\sim OU\text{-}VG(1,5,-0.5,0.015,0.5)$ using the optimal entry levels in the first row of Table \ref{asym}.} \label{asym1}
\end{center}
\end{figure}

The results in Table \ref{asym} show how $d_+^*$ and $d_-^*$ may differ due to skewness.  For negatively skewed spreads, the trade is more likely to enter from the negative level, and so the positive level is slightly lower to compensate. Figures \ref{asym1} and \ref{fig:trade}  show the sample paths of the spread when $\mu=-0.5$ (large negative skew) and  $\mu=0$ (no skew), respectively.

In the classical OU spread model, which has no skew, \cite{leung2015optimal} derive the optimal entry and exit levels with transaction costs, where the two levels can also be asymmetric.

\subsection{Exit Level Different from the Mean}

\begin{figure}[!htb]
    \begin{center}
        \includegraphics[width=.8\textwidth]{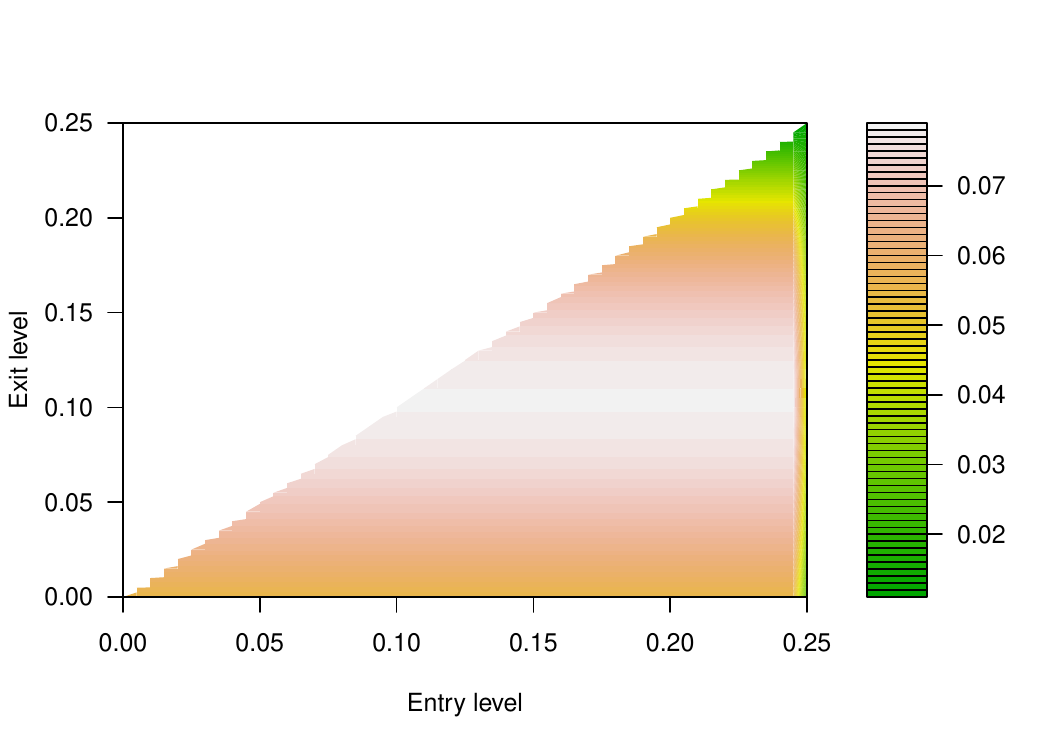}
        \caption{The estimate of the value function $V(c,d)$ with entry level $d$ and exit level $c$. The parameters are $(\lambda,b,\mu,\sigma^2,\eta)=(1,5,0,0.015,0)$, $X_0=0.25$, $\gamma =0$, $r=1$.}
         \label{diff-exit-fig}
    \end{center} 
\end{figure}

Recall that we set $c=c_+=c_-$ for simplicity. So far, we have set $c=0$. However, it is possible to find examples where $c^*$ is much greater than 0. Such a situation can occur when $X_0>\overline{\mu}$ and the discount rate $r$ is very large. Let $X_0=0.25$ and $r=1$ while the other parameters are as before with $(\lambda,b,\mu,\sigma^2,\eta)=(1,5,0,0.015,0)$. We optimize $c,d$ for $0<c<d$, and find that the optimal exit level is $c^*=0.105$, while the optimal entry level $d^*$ is any value $c^*<d^*<0.25$ as shown in Figure \ref{diff-exit-fig}, which plots the estimate of the value function $V(c,d)$.

Thus, the pairs trading strategy is to enter the trade immediately, and because the large discount rate greatly reduces the profit if the trade is not exited quickly, the trade is exited at $c^*=0.105$ above the stationary mean $\overline{\mu}$ rather than waiting for it to fall to $\overline{\mu}$.

\begin{figure}[!htb]
    \begin{center}
        \includegraphics[width=.8\textwidth]{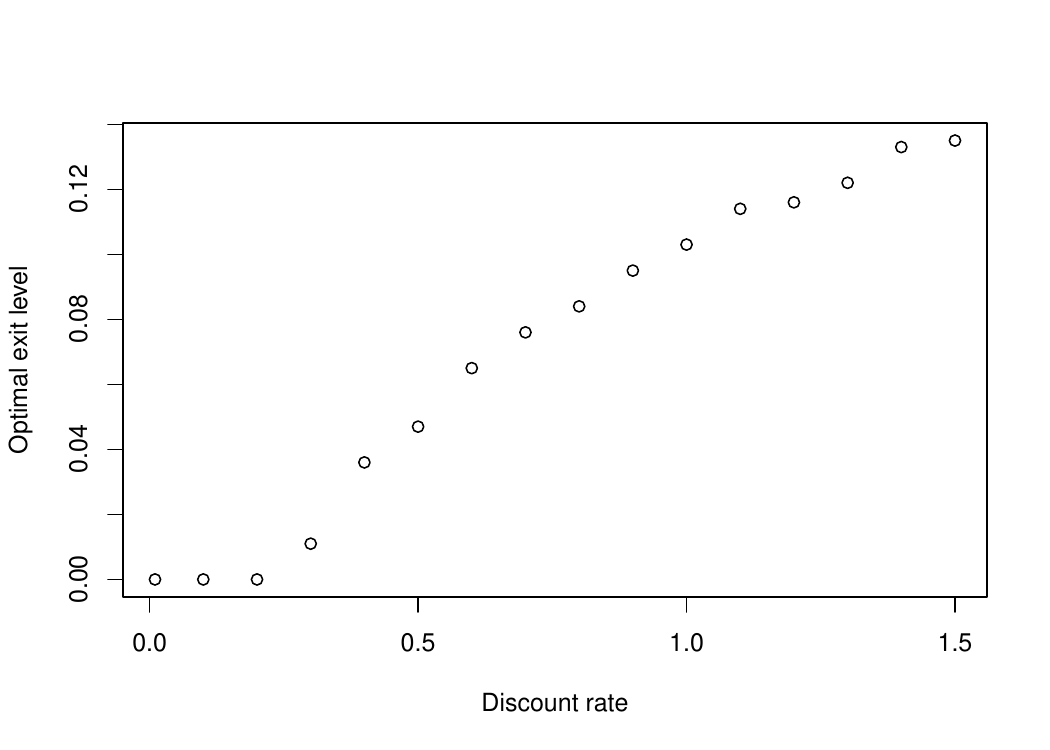}
        \caption{The optimal exit level  $c^*$  for various value of discount rate  $r = 0.01, 0.1,0.2,\dots, 1.5$.  The other parameters are $(\lambda,b,\mu,\sigma^2,\eta)=(1,5,0,0.015, \allowbreak 0)$, $X_0=0.25$, $\gamma =0$.}
        \label{diff-exit}
    \end{center} 
\end{figure}

Figure  \ref{diff-exit} shows how the optimal exit level $c^*$ changes for different values of the discount rate $r$. For the plotted values, the point at which $c^*> 0.000$ (to 3 decimal places) is $r=0.3$.  For all these values of $r$, if the initial value is instead set to $X_0=\overline\mu=0$, then the optimal exit level would be $c^*=0.000$. This demonstrates how having an initial value different from the stationary mean and a large discount rate can make the optimal exit level $c^*$ different from 0. It is difficult to find other situations where this occurs, which suggests optimizing $c$ may have little benefit beyond this, and fixing $c=0$, which corresponds to exiting the trade when the spread passes $\overline{\mu}$, simplifies the optimization.

\subsection{Effect of Jumps}

Now we consider the effect of the parameter $b$, which affects the jumps of the LDOUP $X$, on the optimal entry level $d^*$. As before, let $(\lambda,\mu,\sigma^2,\eta)=(1,0,0.015,0)$.  As $b\to\infty$, the OU-VG process for the spread converges in law to a classical OU process driven by Brownian motion, and the sample paths are closer to continuous. As $b\to0$, roughly speaking, the sample paths become more jumpy (specifically, for each $t$, the BDLP $Z(t)$ converges in distribution to a constant while its kurtosis diverges). Define the overshoot for the entry level $d$ as the random variable  
\begin{align*}
O = \begin{cases}
    |X(\tau_d)-(\overline\mu + d)| & \text{if $\tau_d< \tau_{-d}$},  \\
    |X(\tau_{-d})-(\overline\mu - d)| &  \text{if $\tau_d\geq  \tau_{-d}$.}
\end{cases}
\end{align*}

\begin{table}[htb]
\begin{center}
    \begin{tabular}{c|cccc}   
        \hline
    $b$ & $d^*$ & $\wh V_M(d^*)$  & Sample mean $O$ & Sample sd $O$ \\ \hline\hline
        1 & 0.246 & 0.286 & 0.0640 &  0.0682 \\
        5 &  0.220 &  0.227 & 0.0282 &  0.0306\\
        100 & 0.211 & 0.196 & 0.0086 & 0.0086 \\ \hline
    \end{tabular}
    \caption{For various values of $b$, the optimal entry level $d$, the estimate $\wh V_M(d^*)$ of the optimal expected profit, and overshoot statistics are shown. The other parameters are $(\lambda,\mu,\sigma^2,\eta)=(1,0,0.015,0)$, $X_0=0$, $\gamma =0$, $r=0.01$.}\label{jump-table}
\end{center}
\end{table}

\begin{figure}[!htb]
\begin{center}(a)

    \includegraphics[width=0.8\textwidth,keepaspectratio,trim=0 0cm 0cm 1.5cm,clip]{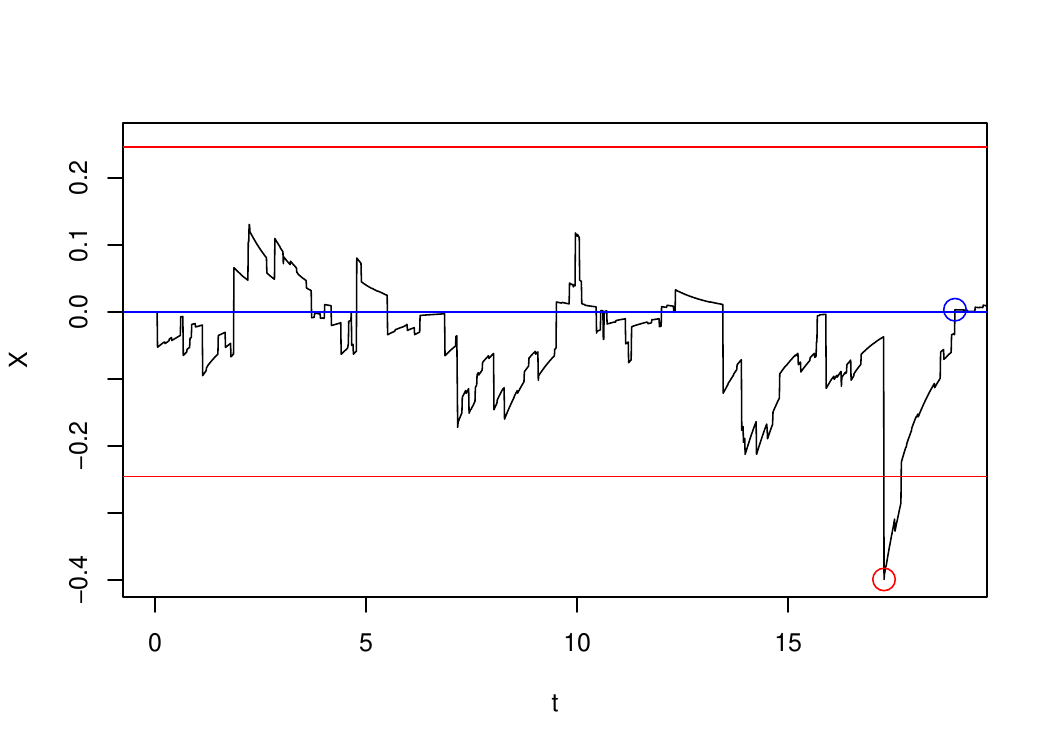}
    
    (b)
    
    \includegraphics[width=0.8\textwidth,keepaspectratio,trim=0 0cm 0cm 1.5cm,clip]{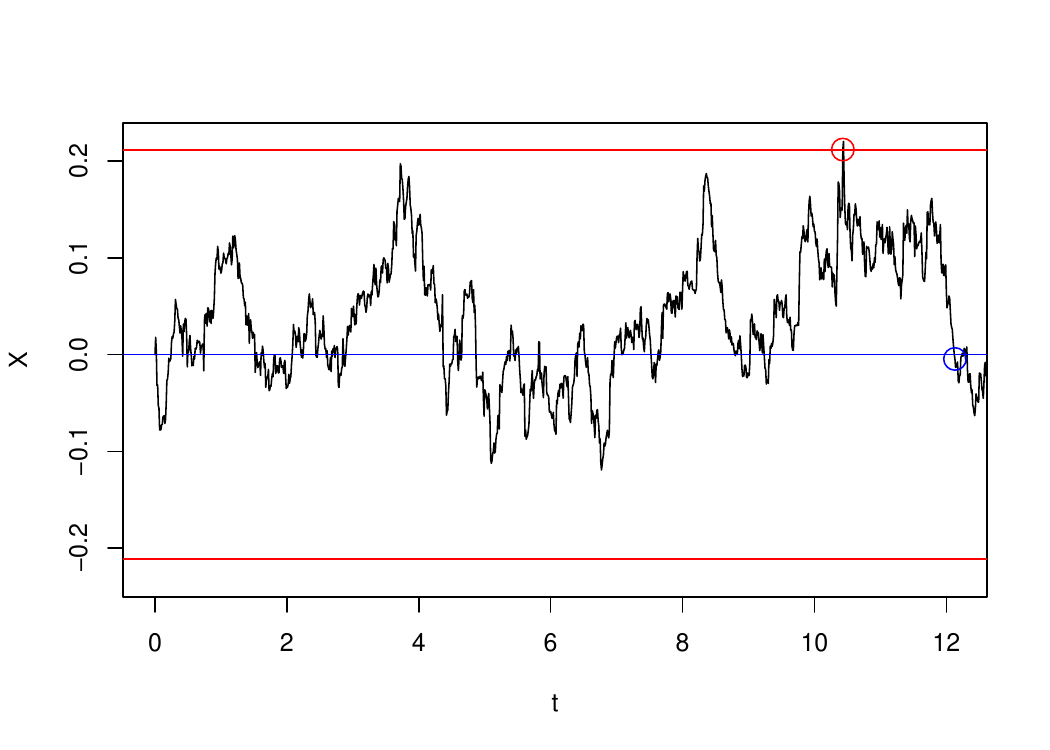}
    \caption{A sample path of the spread  $X\sim OU\text{-}VG(1,b,0,0.015,0)$, where (a) $b=1$ and (b) $b=100$, using the optimal entry levels in the first and third row of Table \ref{jump-table}.}\label{jumpplot}
\end{center}
\end{figure}

Table \ref{jump-table} shows the effect of the sample paths of the spread, roughly speaking, being very jumpy ($b=1$), moderately jumpy ($b=5$), and approximately continuous $(b=100)$. A plot of a sample path in each of these cases is given in Figures \ref{jumpplot}(a), \ref{fig:trade} and \ref{jumpplot}(b), respectively. The results show that the more jumpy the sample path, the higher the level at which the trade is entered to compensate for overshooting. When $b=1$, the mean overshoot is relatively large (0.0640) compared to the optimal entry level $d^*$ (0.246). This demonstrates the importance of accounting for jumps by using LDOUPs compared to the classical OU process. Both Figure \ref{jumpplot} and Table  \ref{jump-table}  show that as the sample path becomes closer to continuous, the mean of the overshoots becomes smaller and the optimal expected profit  $\wh V_M(d^*)$  becomes close to $d^*$ as expected, although they do not converge since $r$ is nonzero.  In this example,   $\wh V_M(d^*)$ increases as $b$ decreases. Note that the stationary variance remains constant, so the increased profitability is not due to changes in the variance.

\section{Bivariate Pairs Trading}\label{sect-two-pairs}
We now consider pairs trading in a bivariate setting, where the spread $\bfX=(X_1,X_2)$ is a bivariate LDOUP with stationary mean $\overline{\bfmu}=(\overline\mu_1,\overline\mu_2)$.  While we focus on the 2-dimensional case, the concepts here can in principle be extended to the $n$-dimensional case.

\subsection{Paris Trading Problem}

While there are many ways to set up a pairs trading strategy on multiple spreads, we use the following setup to generalize the notion of one trade cycle used in the univariate setting. A trade is entered for the first time that either spread $X_k$ is above  $\overline\mu_k + d_k$ or below $\overline\mu_k - d_k$ for $k=1,2$. If a trade on spread $X_k$ is entered, then exit occurs when the spread passes $\overline\mu_k + c_k$ or $\overline\mu_k - c_k$, respectively. When a trade is entered, no further trades can be initiated, even if the other spread later passes the entry level.

More generally, it is possible to have asymmetric entry and exit levels for both spreads, where ${d_{+,1}}$, $d_{-,1}$ are the levels for entering a trade on spread $X_1$ from the positive and negative sides, respectively, and similarly for the other entry and exit levels. For each spread $k=1,2$, there are 8 relevant passage times  defined by
\begin{align*}
    \tau_{d_{+,k}}&=\inf\{t\geq 0 : X_k(t)>\overline \mu_k + d_{+,k}\}\wedge T,\\
    \tau_{c_{+,k}}&=\inf\{t\geq \tau_{d_{+,k}} : X_k(t)< \overline \mu_k + c_{+,k}\}\wedge T,\\
    \tau_{d_{-,k}}&=\inf\{t\geq 0 : X_k(t)< \overline \mu_k - d_{-,k}\}\wedge T,\\
    \tau_{c_{-,k}}&=\inf\{t\geq \tau_{d_{-,k}} : X_k(t)> \overline \mu_k - c_{-,k}\}\wedge T,
\end{align*}
with $0\leq c_{\pm,k}<d_{\pm,k}$. However, we will not need to use this much generality in the examples below.

The nonzero probability of common jumps for the WVAG process used as the BDLP, and hence for $\bfX$, means that the event $\{\tau_{d_{+,1}} = \tau_{d_{+,2}}\}$, for instance, also has nonzero probability, that is,  both spreads can pass their corresponding entry levels simultaneously.  In this case, we invest with 50\% weight in each of the spreads. An alternative approach would be to invest fully in the spread furthest from the mean-reverting level. However, in the numerical examples we consider below, both approaches produce virtually the same results, since it is rare that both spreads pass the optimal entry levels simultaneously. Therefore, we will only consider the former approach.

Let $\tau = \min\{\tau_{d_{+,1}},\tau_{d_{-,1}},\tau_{d_{+,2}},\tau_{d_{-,2}}\}$, $\bfc=(c_{+,1},c_{-,1} , c_{+,2},c_{-,2})$ and $\bfd=(d_{+,1},d_{-,1} , d_{+,2},d_{-,2})$. The profit over one trade cycle is given by
\begin{align*}
   P:= P(\bfc,\bfd) ={}& \eins_{\{\tau \in \{d_{+,1},d_{+,2}\} \}  } \sum_{k=1,2} e^{-r\tau_{c_{+,k}} }   W_k(X_k(\tau_{d_{+,k}}) - X_k(\tau_{c_{+,k}}))\\
      & +\eins_ {\{\tau \in \{d_{-,1},d_{-,2}\} \}  } \sum_{k=1,2} e^{-r\tau_{c_{-,k}} }  W_k (X_k(\tau_{c_{-,k}})-X_k(\tau_{d_{-,k}})  ) ,
\end{align*}
where the weights are
\begin{align*}
    (W_1,W_2) = \begin{cases} (1,0) & \text{if $\tau_{d_{+,1}}<\tau_{d_{+,2}}$ or $\tau_{d_{-,1}}<\tau_{d_{-,2}}$,} \\
        (0,1) & \text{if $\tau_{d_{+,1}}>\tau_{d_{+,2}}$ or $\tau_{d_{-,1}}>\tau_{d_{-,2}}$,} \\
           (0.5,0.5) &  \text{otherwise,} \\
    \end{cases}
\end{align*}
and $r$ is the discount rate.

The objective function is the value function
\begin{align*}
V(\bfc,\bfd)=\EE[P]-\gamma\myVar(P),\quad \gamma\geq 0.
\end{align*}

We want to find the optimal level $\bfc,\bfd$ for the pairs trading strategy to maximize the value function, that is
\begin{align*}
\max_{\bfc,\bfd} V(\bfc,\bfd).
\end{align*}

\subsection{Simulating the Spread}
Now we specifically let the spread be $\bfX\sim OU\text{-}WVAG^2(a,\bfalpha, \bfmu,\Sigma,\bfeta)$. To simulate $\bfX(t_0),\dots,\bfX(t_q)$, we use the Euler scheme approximation in \cite[Sections 4.2 and 5.1]{Lu21}, which is based on taking a stochastic integral representation of $\bfZ^*(\Delta)$, which can be approximately simulated whenever $\bfZ(t)$, in this case using \eqref{wvagpropb}. This is in contrast to the univariate OU-VG process which was simulated exactly in Section \ref{sect-MC} with a method that has no multivariate generalization. We now determine the optimal trading levels using  Monte Carlo methods.

\subsection{Numerical Implementation and Analysis}

We consider two examples of bivariate pairs trading on the spread $\bfX\sim OU\text{-}\allowbreak WVAG^2(\lambda,a, \bfalpha,\allowbreak\bfmu,\Sigma,\bfeta)$ simulated using the approximate method above with stepsizes $\Delta= 0.01$ and $\wt \Delta=0.001$, and terminal time $T=50$.  In all cases, the initial value is $\bfX_0=\bfnull$, $\gamma=0$, we do not consider control variates, and  $\bfmu=-\bfeta$ so that  $\overline{\bfmu}=\bfnull$. Unless stated otherwise, we assume for simplicity that the entry levels are symmetric with $d_k=d_{+,k}=d_{-,k}$, $k=1,2$, and the exit levels are $\bfc=\bfnull$, so we optimize the value function over $d_1,d_2$. We use $m=10,000$ Monte Carlo simulations. 

In our numerical examples, we explore the effect of correlation on the optimal entry level and the optimal expected profit. In the first example, the correlation has little effect, whereas in the second example, it has a larger effect.  The correlation of the components of $\bfX$ is in part controlled by the correlation parameter $\rho= \Sigma_{12}/\sqrt{\Sigma_{11}\Sigma_{22}}$. However, note that $\rho$ is the correlation of the Brownian motion subordinate in \eqref{ws}, and is related to the correlation of the components of $\bfX$ by
\begin{align}
\myCorr(X_1(t),X_2(t))&=\myCorr(Z_1(t),Z_2(t))\nonumber\\
&=\frac{a((\alpha_1\wedge \alpha_2)\Sigma_{12}+\alpha_1\alpha_2\mu_1\mu_2)}{\sqrt{\Sigma_{11}+\alpha_1\mu_1^2}\sqrt{\Sigma_{22}+\alpha_2\mu_2^2}}\label{coreqn}
\end{align}
for all $t>0$,   due to  \eqref{covx} and \eqref{cv2}.

\paragraph{Example 3.} Let 
\begin{align}
\lambda = 1,\; a = 2.5, \; \bfalpha =\begin{pmatrix}0.2\\0.3\end{pmatrix}, \; \bfmu =\begin{pmatrix}0\\-0.2\end{pmatrix}, \; \Sigma &= \begin{pmatrix}0.015 & \rho \sqrt{\Sigma_{11}\Sigma_{22}}\\{*}&   0.02 \end{pmatrix}, \label{ouwpar}
\end{align}
and $r=0.01$.

\begin{table}[htb]
    \begin{center}
        \begin{tabular}{c|ccccc}
            \hline
            $\rho$ & $0.9$ & $0.3$ & 0 & $-0.3$ & $-0.9$ \\
            $\myCorr(X_1(t),X_2(t))$ & $ 0.356$ & $0.119 $ & 0 & $-0.119 $ & $- 0.356$ \\
            \hline         \hline
            $\wh V_M(d_1^*,d_2^*)$   & 0.334 & 0.335 &  0.336 & 0.336 &  0.334\\
            $d_1^*$ &  0.318 & 0.311 & 0.305 & 0.316  &  0.316\\
            $d_2^*$&  0.338 & 0.345 & 0.343  & 0.334 &0.345\\
            \hline  
           Trades only $X_1$ (\%) &  11.42 & 14.15    &15.88  & 12.72  & 12.40\\
           Trades only $X_2$ (\%) &  83.08 &  80.34 &  79.46  & 83.15   & 81.38\\
           Trades both (\%) &        0.97 & 0.76  & 0.58 & 0.56      & 0.74\\
           Trades neither (\%)    &  4.53 & 4.75   & 4.08   & 3.57 &    5.48\\
           \hline
        \end{tabular} 
        \caption{For various values of $\rho$, the optimal entry levels $d_1^*, d_2^*$, the estimate  $\wh V_M(d_1^*,d_2^*)$ of the optimal expected profit, and the estimated probability of which spread $X_1,X_2$ is traded are shown. The other parameters are  given in \eqref{ouwpar} with $\bfX_0=\bfnull$, $\gamma=0$, $r=0.01$.}\label{multi-table}
    \end{center}
\end{table}

Table \ref{multi-table} shows the optimal entry levels $d_1^*$ and $d_2^*$, the optimal expected profit $\wh V_M(d_1^*,d_2^*)$, and the estimated probability of which spread is traded, and each of these quantities is approximately equal across the values of $\rho$ considered. 

When pairs trading on the univariate spreads $X_1$ and $X_2$, the optimal expected profit is 0.227 (see the second row of  Table \ref{jump-table}) and 0.331 respectively, while it is  0.334--0.336 in this bivariate pairs trading example. Bivariate pairs trading is more profitable than univariate pairs trading on either spread, although it is approximately as profitable as trading only $X_2$ since this accounts for 79--84\% of trades.  The optimal expected profit of the bivariate pairs trade must be at least that of the best univariate pairs trade, as the latter strategy can be implemented within bivariate pairs trading by setting the other spread to have an arbitrarily large optimal entry level. Indeed, this is happening to some extent here as the optimal entry level of the spread $X_1$ has increased from 0.220 in the univariate case to  0.305--0.318  in the bivariate case. Thus, this decreases the probability of only $X_1$ entering a trade from 96.79\% to 11--16\% in Table  \ref{multi-table} (the latter accounts for the probability that a trade on $X_1$ fails to be entered because a trade on $X_2$ is entered first, while the former is univariate pairs trading and does not include this). So in this example, the method has essentially identified that $X_2$ is the more profitable spread, regardless of correlation, and has accordingly increased the optimal entry level of $X_1$ to reduce the instances where it is traded.

\paragraph{Example 4.} Now we consider an example where different values of the correlation parameter $\rho$ can have a larger effect on the optimal entry level and optimal expected profit. Suppose the parameters are now
\begin{align}
\lambda = 1,\quad a = 6.65,\quad\bfalpha =\begin{pmatrix}0.15\\0.15\end{pmatrix}, \quad \bfmu =\bfnull, \quad \Sigma = \begin{pmatrix}0.015 & \rho \sqrt{\Sigma_{11}\Sigma_{22}}\\ {*}  & 0.015 \end{pmatrix}, \label{ouwpar2}
\end{align}
and $r=1$.

Recall the parameter boundary $a<\frac{1}{\alpha_1}\wedge \frac{1}{\alpha_2}$. In Example 3, the upper bound of $\myCorr(X_1(t),X_2(t)) $ as a function of $\rho\in(-1,1)$  is  $0.395$ by  \eqref{coreqn}. The upper bound converges to  1 as   $a\to  \frac{1}{\alpha_k}$ while keeping $\alpha_k,\mu_k,\Sigma_{kk},\eta_k$  constant for all $k$. Since $a \approx \frac{1}{\alpha_1}=\frac{1}{\alpha_2}$ here, this example represents monitoring two spreads  $X_1\eqd X_2$ that are equal in law for each fixed $\rho$, with $X_1$ and $X_2$ converging pathwisely as $\rho\to1$, since in the limit, $\myCorr(X_1(t),X_2(t))=1$ for all $t$ with $\bfX_0=\bfnull$. Also, appealing to the symmetry of $\bfX(t)$ for each fixed $t$, all components of the optimal entry levels $\bfd^*$ are equal to  $d^*$, say. Thus, we optimize over $d$, the symmetric entry level for all components.

\begin{table}[htb]
    \begin{center}
        \begin{tabular}{c|cccccc}
            \hline
            $\rho$ & $0.99$ & $0.9$ & 0.3 & 0 & $-0.3$ & $-0.9$ \\
            $\myCorr(X_1(t),X_2(t))$ & 0.988 &  0.898 & 0.299 & 0 &  $-0.299$ & $-0.898$\\
            \hline\hline 
            $\wh V_M(d^*)$  & 0.032  &0.035&    0.039 &  0.040 &  0.040  &   0.034\\
            $d^*$ & 0.037  & 0.044 &   0.045 &  0.045 &    0.044  &  0.045\\
   \hline
        \end{tabular} 
    \caption{For various value of $\rho$, the optimal entry level $d^*$ and the estimate  $\wh V_M(d^*)$ of the optimal expected profit. The other parameters are  given in \eqref{ouwpar2} with $\bfX_0=\bfnull$, $\gamma=0$, $r=1$.} \label{ex2table}
    \end{center}
\end{table}

Table \ref{ex2table} shows the results for this example over different values of $\rho$.
 Here, the optimal expected profit $\wh V_M(d^*)$ increases from 0.032 to 0.040 as $\rho$ decreases from 0.99 to 0, this represents a 26\% increase in the optimal expected profit. The case where $\rho =0.99$ is approximately the case of univariate pairs trading, which occurs in the limit as $\rho\to1$. Since  $\wh V_M(d^*)$  is larger when $\rho$ is much less than 1, the results demonstrate a  situation where bivariate pairs trading is much more profitable than what is effectively pairs trading on a single spread and where correlation has a larger effect.

These results are consistent with intuition. Monitoring multiple spreads of the same law presents additional opportunities to enter into the trade sooner, which increases the optimal expected profit relative to monitoring only one spread while having a larger discount rate $r$ substantially rewards entering the trade sooner. Thus, in this example, using bivariate pairs trading by choosing pairs that are uncorrelated (as represented by the $\rho=0$ case)  is much more profitable than either univariate pairs trading or choosing highly correlated spreads (both represented by the $\rho\to1$ case).

\section{Conclusions}\label{sect-conclude}
We have presented a Monte Carlo framework for evaluating pairs trading strategies where the mean reverting spread follows an LDOUP. We numerically demonstrated how different parameters affect the optimal trading level and value function and capture various aspects of the trading strategy.  The framework is flexible to accommodate a wide class of LDOUPs, provided that they can be fitted and simulated, as well as different ways of trading the spread.  For instance, methods for fitting and simulating univariate LDOUPs with tempered stable and normal inverse Gaussian stationary distributions are provided in \cite{vstt}. Our proposed control variates for variance reduction, can also be applied to similar problems.

For future research, one useful direction is to incorporate transaction costs into the trading problems. To that end, \cite{do2012pairs} examine the profitability of pairs trading accounting for transaction costs, and \cite{leung2015optimal} analyze an optimal stopping approach to pairs trading with transaction costs and stop loss, and this work could be extended to spreads following LDOUPs.   Another possibility is to consider large jumps in stock prices that trigger immediate liquidation due to the need to return the stocks. This can also be incorporated into the stochastic model and value function. Additional extensions of our framework include portfolio optimization problems arising from trading multiple spreads and spread selection based on statistical and other characteristics. Theoretical results on pairs trading under LDOUP models are very limited, and additional work in this direction would be useful. However with LDOUPs, these problems are complicated and unlikely to admit analytic solutions, so simulation-based approaches can be both mathematically interesting and practically useful.

\appendix \section{Equivalence of control variates and linear regression} \label{appendix}

Suppose we have multiple control variate $\bfC = (C_1,\dots, C_p)$ to estimate $\EE[P]$, where $(\bfC,P)$ is jointly simulated as we want the control variates to be correlated with $P$. As explained in \cite[Section 4.1.2]{Glas}, the control variate estimator is
\begin{align*}
    \wh P_{C} = \overline{Y} - \wh\bfbeta_C'(\overline{\bfX}-\bfmu_C),
\end{align*}
where $\bfY$ is the vector of $m$ simulated values of $P$, $\overline{Y}$ is its sample mean, $\bfX_1,\dots, \bfX_p$ are the vectors of  $m$ simulated values of $C_1,\dots,C_p$, respectively, $\overline{\bfX}$ is the vector of the sample means of $\bfX_1,\dots, \bfX_p$,  $\bfmu_C = \EE[\bfC]$, $\wh\bfbeta_C = S_{XX}^{-1}S_{XY}$,  $S_{XX} \in\RR^{p\times p}$ with the sample covariance $ \wh{\myCov}(\bfX_i,\bfX_j)$ as its $(i,j)$-entry, and  $S_{XY} \in\RR^{p}$ with the sample covariance $ \wh{\myCov}(\bfX_i,\bfY)$ as its $i$th entry. This generalizes the method of Section \ref{cv-sec}.

Consider the linear regression model $\bfY = X\bfbeta +\bfepsilon$, where $X \in \RR^{(p+1)\times m}$ is the design matrix, which includes an intercept and the control variates $C_1,\dots, C_p$, and $\bfbeta\in\RR^{p+1}$ with LSE $\wh\bfbeta$. The prediction of this linear regression  at the point  $\bfx = (1,\bfmu_C)$ is
\begin{align*}
    \wh Y=   \wh\bfbeta'\bfx  = \overline{Y} + \wh\bfbeta_C'(\bfmu_C-\overline{\bfX}) =   \wh P_{C},
\end{align*}
where the second equality follows from the centered form of linear regression by \cite[pg 156, Equations (7.36), (7.46)]{ReSc08}. This reference also provides that  $ \wh\bfbeta_1 =  \wh\bfbeta_C$, where $\wh\bfbeta = (\wh\beta_0 , \wh\bfbeta_1)$.

Thus, the control variate estimator $\wh P_C$ is the prediction of this linear regression. This fact is well-known and mentioned in  \cite[Section 4.1.1]{Glas} for a single control variate, and it is unsurprising that it is also true for multiple control variates. \cite{Nel90} gives a different but equivalent statement where the regressors are $\bfC-\bfmu_C$.

\subsection*{Acknowledgments}
Kevin Lu thanks the Department of Applied Mathematics at the University of Washington, where the substantial majority of his work on this article was completed. We thank an anonymous referee for the useful comments which have helped improve the paper.

\subsection*{Disclosure Statement}

The authors report there are no competing interests to declare.

\bibliographystyle{apalike}
\bibliography{bibliography}   
\end{document}